\newcommand{\be}{\begin{equation}}\newcommand{\ee}{\end{equation}}
\newcommand{\bea}{\begin{eqnarray}}\newcommand{\eea}{\end{eqnarray}}
\documentstyle[12pt]{article}
\def\a{\alpha}\def\b{\beta}\def\g{\gamma}\def\d{\delta}\def\e{\epsilon}
\def\k{\kappa}\def\l{\lambda}\def\L{\Lambda}
\def\Th{\Theta}\def\th{\theta}

\newcommand{\p}[1]{(\ref{#1})}
\begin{document}
\renewcommand{\thefootnote}{\fnsymbol{footnote}}
\thispagestyle{empty}
\begin{center}
{\large \bf  ON ANOTHER VERSION \\
 OF THE TWISTOR--LIKE APPROACH \\
TO  SUPERPARTICLES }
\vspace{1.5cm}
\footnote{Work supported in part by the
International Science Foundation under the Grant RY No 9000,
and by Ukrainian State Committee on Science and Technology under the
contract No 2.3/664.
}
\\ Igor A. Bandos,
\\ Aleksey Yu. Nurmagambetov
\\ Dmitrij P. Sorokin,
\\ and
\\ Dmitrij V. Volkov
\vspace{1cm}\\
\renewcommand{\thefootnote}{\dagger}
{\it Kharkov Institute of Physics and Technology}\\
{\it 310108, Kharkov, the Ukraine}\\
e-mail:  kfti@kfti.kharkov.ua\\

\vspace{1.5cm}
{\bf Abstract}

\vspace{1cm}

Considered is a worldsheet supersymmetric generalization of the
$D=3$ Ferber--Schirafuji twistor--superparticle action.
\end{center}

\vspace{1cm}\\
{\bf hep-th /9409...}; \\
{\sl Submitted to JETP.Lett.}
\bigskip
\bigskip \\
{\bf September 1994}
\setcounter{page}1
\renewcommand{\thefootnote}{\arabic{footnote}}
\setcounter{footnote}0
\newpage

Twistor--like doubly supersymmetric formulations of superparticles
\cite{stv,vz,stvz,sp,ps,gs92}, superstrings \cite{hsstr,hsstr2,gs93} and
supermembranes \cite{tp93} have attracted great deal of attention, in
particular, because of a hope to break through the long--standing problem
of the covariant quantization of these theories.

In the twistor--like approach the infinite--reducible fermionic
$\k$--symmetry \cite{al,sig}, which causes the problem of covariant
quantization \cite{gsw}, is replaced by local worldsheet supersymmetry
which is irreducible by definition, and the theory is formulated as a
superfield theory in a worldsheet superspace imbedded into a space--time
target superspace. So that the model of this kind possesses doubly
supersymmetry.

Earlier doubly supersymmetric dynamical systems (of more general physical
contents) were considered by several groups of authors irrelative to the
$\k$--symmetry problem \cite{spinsup}.

Several versions of twistor--like doubly supersymmetric particles and
heterotic strings have been constructed in
 $D=3,4$ and $6$ dimensions of space--time, while in D=10 only one
superfield formulation is known \cite{gs92} and, unfortunately, the latter
itself suffers the infinite reducibility problem arising for a new local
symmetry \cite{gs92} being crucial for the possibility of eliminating
auxiliary degrees of freedom of the objects under consideration
\footnote{
Note, however, that at the component level, when auxiliary
fields were eliminated by gauge fixing and solving for relevant equations
of motion, all remaining local symmetries are irreducible. This also takes
place in a twistor--like Lorentz--harmonic formulation of
super--$p$--branes \cite{bh,bzm} developed in parallel to
the superfield twistor approach.
}.

So the main motivation of the present paper is, from the one-hand side, to
develop a version of the twistor--like formulation which would be free of
the reducibility problem already at the superfield level, and, from the
other hand, would look ``twistor--like'' as much as possible. The letter,
as we hope, may allow one to better utilize the powerful twistor
techniques for deeper implementation of the twistors into the structure of
supersymmetric theories.

The superfield
twistor--like models of $N=1$ Brink--Schwarz superparticles in $D=3,
4, 6$ and $10$ considered so far are based on the doubly supersymmetric
generalization of the following massless bosonic particle action
\cite{stv}:
\begin{equation}\label{1}
S= \int d \tau p_{{m}} ( {\dot x}^{{m}} - \bar\l \g^{{m}} \l ) , \qquad
\end{equation}
where $p_{{m}} $ is the particle momentum and $\l^\a$ is a commuting
spinor variable ensuring the validity of the mass shell condition
$
p_{{m}} p^{{m}} = 0 = {\dot x}_{{m}}  {\dot x}^{{m}}
$
due to the Cartan--Penrose representation
$
{\dot x}^{{m}} = \bar\l \g^{{m}} \l
$
of the light--like vectors in $D=3, 4, 6$ and $10$ space--time dimensions.

The straightforward doubly supersymmetric generalization of \p{1} is
\cite{stv,gs92}

\begin{equation}\label{1.2}
S= \int d \tau d^{{D-2}} \eta
P_{{mq}} ( D_{{q}} X^{{m}} - D_{{q}} \bar{\Theta} \g^{{m}}
\Theta ) , \qquad
\end{equation}
where the number $n = D - 2$ of the local worldline supersymmetries is
equal to the number of the $\k$--symmetries in  $D=3, 4, 6$ and $10$ ;
$
D_{{q}} = {{\partial} \over {\partial \eta^{{q}} } }+i \eta
\partial _{\tau}
$
is an odd supercovariant derivative in a worldline superspace
$(\tau , \eta^{{q}} )$ ,
$
\{ D_{{q}}, D_{{p}} \} = 2i \d_{pq}  \partial _{\tau}
$
and $( X^{{m}}, \Th^{{\a}} )$ being worldline superfields which
parametrize the ``trajectory'' of the superparticle in a target
superspace. Bosonic spinor variables $ \l^{{\a}}_{{q}}$ appear in \p{1.2}
as a superpartners of Grassmann coordinates $ \th^{{\a}} = \Th^{{\a}}
\vert_{\eta = 0} $ :

\begin{equation}\label{1.3}
\l^{{\a}}_{{q}} =
D_{{q}} \Theta^{{\a}} (\tau, \eta) \vert_{\eta = 0}
\qquad
\end{equation}

The analysis of the action \p{1.2} \cite{stv,gs92} shows that it describes
a superparticle classically equivalent to the massless $N=1$
Brink--Schwarz superparticle in $D=3, 4, 6$ and $10$ .

As we have already mentioned, in $D=4, 6$ and $10$ the action \p{1.2}
possesses a local symmetry \cite{gs92} under the following transformations
of the Lagrange multiplier $P_{{mq}}$ :
\begin{equation}\label{1.4}
\d P_{{mq}} = D_{{p}} \bar{\Xi}_{{qpr}}\g_mD_{{r}} \Th ,
\qquad
\end{equation}
with
$ \bar{\Xi}^{{\a}}_{{qpr}}$ being symmetric and traceless with respect to
the indices $(p, q, r)$. This symmetry is infinite reducible since
$P_{{mq}}$ is inert  under the transformations \p{1.4} with
\begin{equation}\label{1.5}
\bar{\Xi}^{{\a}}_{{qpr}}
= D_{{s}} \bar{\Xi}^{{\a}}_{{qprs}}
\qquad
\end{equation}
where $ \bar{\Xi}^{{\a}}_{{qprs}}$ is again symmetric and traceless,
and \p{1.5} is trivial if
$
\bar{\Xi}^{{\a}}_{{qprs}}
= D_{{s}} \bar{\Xi}^{{\a}}_{{qprst}}
$
and so on and so far.

The reducibility of the transformations \p{1.4} is akin to the
reducibility of the gauge symmetries of the antisymmetric gauge fields.
It is just the problem of reducible symmetries in these theories that
stimulated further development the quantization procedure which was
consistently fulfilled for finite reducible symmetries \cite{bv}
to  which the gauge transformations of the antisymmetric bosonic tensor
fields belong to. However, the general receipt for dealing with the
infinite reducible symmetries is still unknown (see \cite{infred} and
refs. therein). Thus, one has to avoid this problem one way or another .
In the case under consideration one may try to find another form of the
twistor--like superfield action for the superparticle.

To this end, let us choose, as a starting point, the form of the
twistor particle action considered by Ferber \cite{fer} and Schirafuji
\cite{schir}
\begin{equation}\label{5}
S= \int d \tau  \bar\l\g_{{m}}\l{\dot x}^{{m}} ,
\end{equation}

For simplicity,  we shall consider the case of $N=1,~D=3$ superparticle.

To generalize \p{5} to the doubly supersymmetric case one could naively try
(using (3)) to wright down an action in the following form
\begin{equation}\label{6}
S= \int d \tau d \eta
D \Theta_{{\a}} D \Theta_{{\b}} D X^{{\a \b}} , \qquad
\end{equation}
Where
$
X^{{\a \b}} \equiv X^{{m}} \g_{{m}}^{{~\a \b}}.
$

However, action \p{6} does not describe $N=1,~D=3$ Brink--Schwarz
superparticle, but a model with an odd physical contents.
The reason is that \p{6} is invariant under
$$
\d \Th^{{\a}} = \e_{{1}}^{\a} , \qquad
\d X^{\a \b} = \Th^{\a} \e^{\b}_{{2}} + \Th^{\b} \e^{\a}_{{2}} ,
\qquad
$$
so, that the target space is not the usual superspace, but one with
additional $\theta$--translation transformations.

Note that action \p{6} is part of a so called spinning superparticle model
considered several years ago \cite{spinsup}.

To construct a doubly supersymmetric action for describing an $N=1$
Brink--Schwarz superparticle we have to keep only one target space
supersymmetry. The right action turns out to be as follows
\begin{equation}\label{7}
S= \int d \tau d \eta
\L_\a \L_\b (D X^{{\a \b}} - i D \Theta^{\a} \Theta^{\b}
- i D \Theta^{\b} \Theta^{\a} ) , \qquad
\end{equation}
where $\L_{\a} ( \tau , \eta )$ is a commuting spinor superfield.

In addition to $N=1$ target space supersymmetry and $n=1$ local worldline
supersymmetry
\begin{equation}\label{10}
\d \eta = {i\over 2} D\Xi(\tau, \eta), \qquad
\d \tau = \Xi + {1\over 2} \eta D\Xi , \qquad
\d D = -{1\over 2} \partial \Xi D
\qquad
\end{equation}
action \p{7} is invariant under bosonic transformations
\begin{equation}\label{9}
\d X^{\a \b} = b(\tau, \eta) \L^{\a} \L^{\b} , \qquad
\d \Th ^{\a} = 0 = \d \L ^\a
\qquad
\end{equation}
and under a superfield  irreducible counterpart of the conventional
fermionic  $\kappa$--symmetry
\begin{equation}\label{8}
\d \Th ^{\a} = \k (\tau, \eta) \L^{\a} , \qquad
\d X^{\a \b} = 2i \d \Th^{\{ \a} \Th^{\b \} } , \qquad
\d \L ^\a = 0 , \qquad
\end{equation}
which resembles the fermionic symmetry of
twistor--like component actions for super--$p$--branes \cite{stv,bzm}
( the braces $\{ ...\}$ denote symmetryzation of the indices).

The algebra of the transformations \p{9}, \p{8} is closed.

The equations of motion derived from \p{7} are
\begin{equation}\label{11a}
\Pi^{\a \b} \L_{\b} \equiv
( D X^{\a \b} - 2i D \Th^{\{ \a} \Th^{ \b \} } ) \L_{\b} = 0 ,
\qquad
\end{equation}
\begin{equation}\label{11b}
\L_{\b} D \Th^{\b} = 0 , \qquad
\end{equation}
\begin{equation}\label{11c}
\L_{\{ \a} D \L_{\b \} } = 0 ,
\qquad
\end{equation}

The general solutions to \p{11a} and \p{11b} are, respectively,
\begin{equation}\label{12a}
\Pi^{\a \b} = \Psi (\tau, \eta) \L^{\a} \L^{\b} , \qquad
\qquad
\end{equation}
\begin{equation}\label{12b}
D \Th^{\a} = a(\tau, \eta) \L^{\a} , \qquad
\end{equation}
At the same time, from \p{11c} it follows that
\begin{equation}\label{12c}
D \L ^\a = 0 , \qquad
\end{equation}

On the mass shell \p{12a} -- \p{12c} the fermionic superfield $\Psi$ and
the bosonic superfield $a$ transform under \p{8}, \p{9} and \p{10} as
follows:
\begin{equation}\label{13}
\d \Psi = D b - {1\over 2} \partial_{\tau} \Xi \Psi
- 2i a \k ,  \qquad
\d a = D \k - {1\over 2}\partial_\tau \Xi a ,
\end{equation}
Hence, one can fix a gauge
\begin{equation}\label{14}
\Psi = 0,  \qquad   a = 1 ,
\end{equation}
at which \p{12a}, \p{12b} are reduced to
\begin{equation}\label{15a}
\Pi^{\a \b} = 0,  \qquad
\end{equation}
\begin{equation}\label{15b}
D\Th^{\a} = \L^{\a},  \qquad
\end{equation}
This gauge
\footnote{ Note, that the gauge choice $a = 0$ in Eq. \p{14} is
inadmissible since  then case from \p{12a}, \p{12b} it would follow that
${d \over d\tau} X^m \vert_{\eta = 0} = 0$,
which is, in general, incompatible with boundary conditions
$X^m (\tau_1) \vert_{\eta = 0} = x_{1}$,
$X^m (\tau_2) \vert_{\eta = 0} = x_{2}$.
}
is conserved  under the $\k$--symmetry reduced to
the worldline supersymmetry
\begin{equation}\label{17}
D\k-{1\over 2}\partial_\tau \Xi = D(\k+{i\over 2}D\Xi) = 0.
\end{equation}
As a result the twistor superfield $\L^{\a}$ is expressed
in terms of $D\Th^{\a}$ and does
not carry independent degrees of freedom, and in the gauge \p{14} the
equations for $X^{\a\b}$ and $\Th^\a$ coincide with those in the
conventional twistor--like formulation \p{1.2} \cite{stv,gs92}.

Thus we conclude that the doubly supersymmetric action \p{7} is
classically equivalent to \p{1.2} and describes the massless $N=1$
superparticle.

The relationship between the two actions can be understood using the
following reasoning. It was proved in \cite{ps} that for $n=1$ action
\p{1.2} is classically equivalent to
\begin{equation}\label{18}
S=\int d\tau d\eta(P_{\a\b}\Pi^{\a\b}-{1\over 2}EP_{\a\b}P^{\a\b})
\end{equation}
due to the existence of the following counterparts of the symmetry
transformations \p{8}, \p{9} \cite{ps}
\begin{equation}\label{19}
\d X^{\a\b}=\tilde bP^{\a\b},\qquad \d E=D\tilde b, \qquad\Th^\a=0,
\end{equation}
\begin{equation}\label{20}
\d X^{\a\b}=2i\d\Th^{\{\a}\Th^{\b\}},\qquad \d E=-2i\k_\a D\Th^\a, \qquad
\d\Th^\a=\k_\b P^{\b\a},
\end{equation}
which allow one to put the Grassmann superfield $E(\tau,\eta)$ equal to
zero globally on the worldline superspace. \footnote{Note that in contrast
to \p{8} the transformations of eq.~\p{20} correspond to an infinite
reducible $\k$--symmetry \cite{al,sig,gsw}.}

At the same time the variation of \p{18} with respect to $E(\tau,\eta)$
leads to the equation
\begin{equation}\label{21}
P^{\a\b}P_{\a\b}=0,
\end{equation}
which can be solved as
\begin{equation}\label{22}
P_{\a\b}=\L_\a\L_\b
\end{equation}
with $\L_\a$ being an arbitrary bosonic spinor superfield. Substituting
\p{22} back into \p{18} we  get just the action \p{7}.

In conclusion we have constructed a version of the twistor--like
formulation of the massless $N=1$, $D=3$ superparticle based on eq.~\p{7}
with all symmetries of the model being irreducible. Action \p{7} looks
very much like a worldline superfield generalization of the supertwistor
action by Ferber \cite{fer}.

One can even rewrite (8) in a complete supertwistor form
of Schirafuji \cite{schir} by introducing the second bosonic spinor
component and Grassmann component of supertwistor \cite{fer}:
\begin{equation}\label{28}
M^{\a} =
X^{\a \b} \L_{\b}, \qquad \Upsilon = \Theta^{\a} \L_{\a} \qquad
\end{equation}
Then with taking into account constraint \p{28} action (8) takes the form
$$
S= \int d \tau d \eta (\L_{\a} D M^{\a}
- D \L_{\a} M^{\a}
- 2i \Upsilon D \Upsilon ) \qquad
$$

Note that the transformations \p{8} resemble
an extra hidden local worldline supersymmetry which relates $\Th^\a$ and
$\L^\a$, and it would be of interest to understand the nature and the role
of this symmetry in more detail.

Action \p{7} admits a generalization to D=4 and 6 superparticles and,
possibly, heterotic strings within the line of the twistor--like
formulation developed in \cite{stv,sp,ps,hsstr2}, where the notion of a
doubly Grassmann analiticity (for D=4) and a doubly harmonic analiticity
(for D=6) have been explored. But the generalization to the case of D=10
twistor--like objects with irreducible local symmetries seems to be more
subtle.  Work on this subject is in progress.\\
\vspace{0.5cm}\\
{\sl \bf Acknowledgments}: The authors are thankful to A. Pashnev for
discussion.

  \end{document}